\documentclass[11pt]{article}
\newcount\biblio
\title{Aspects of Lagrange's Mechanics and their legacy}
\author{\small\textsc{Giovanni Gallavotti}
\\
\small Accademia dei Lincei \& INFN-Roma1,
Italia}%

\usepackage{ifthen}
\usepackage{hyperref}
\usepackage{fancyhdr}
\let\a=\alpha \let\b=\beta  \let\g=\gamma  \let\d=\delta \let\e=\varepsilon
  \let\h=\eta   \let\th=\theta \let\k=\kappa \let\l=\lambda
\let\m=\mu    \let\n=\nu    \let\x=\xi     \let\p=\pi    
 \let\t=\tau   \let\f=\varphi 
  \let\ps=\psi  \let\o=\omega
   \let\L=\Lambda

\let\wt=\widetilde

\let\0=\noindent
\def\defi{\,{\buildrel def\over=}\,}
\def\CC{{\cal C}}
\def\MM{{\cal M}}\def\BB{{\cal B}}
\def\HH{{\cal H}}
\def\otto{\,{\kern-1.truept\leftarrow\kern-5.truept\to\kern-1.truept}\,}

\def\*{\vskip2mm}
\def\lis#1{{\overline#1}}
\let\dpr=\partial
\def\Ie{{\it i.e.\ }}  \def\eg{{\it e.g.\ }}
\def\Eq#1{{\label{#1}}%
\write15{\string\Fe{\string#1}{\ref{#1}}}}
\def\equ#1{(\ref{#1})}

\def\be{\begin{equation}}\def\ee{\end{equation}}
\def\tende#1{\,\vtop{\ialign{##\crcr\rightarrowfill\crcr
 \noalign{\kern-1pt\nointerlineskip} \hskip3.pt${\scriptstyle
 #1}$\hskip3.pt\crcr}}\,}
\def\V#1{{\bf#1}}

\def\iniz{\setcounter{equation}{0}
\rhead{\thepage}\lhead{{{{\small\bf\thesection:}\ \small\SEC}}}
}

\def\eqalign#1{\null\,\vcenter{\openup\jot
  \ialign{\strut\hfil$\displaystyle{##}$&$\displaystyle{{}##}$\hfil
      \crcr#1\crcr}}\,}
\def\bea{\begin{eqnarray}}
\def\eea{\end{eqnarray}}

\newdimen\xshift \newdimen\xwidth \newdimen\yshift \newdimen\ywidth
\def\ins#1#2#3{\vbox to0pt{\kern-#2pt\hbox{\kern#1pt #3}\vss}\nointerlineskip}
\def\eqfig#1#2#3#4#5{
\par\xwidth=#1pt \xshift=\hsize \advance\xshift
by-\xwidth \divide\xshift by 2
\yshift=#2pt \divide\yshift by 2
{\hglue\xshift \vbox to #2pt{\vfil
#3 \includegraphics{#4.eps}
}\hfill\raise\yshift\hbox{#5}}}
\def\8{\write12}  

\font\tenmib=cmmib10 
\font\sevenmib=cmmib7\font\fivemib=cmmib5
\mathchardef\Ba   = "050B  
\mathchardef\Bb   = "050C  
\mathchardef\Bg   = "050D  
\mathchardef\Bd   = "050E  
\mathchardef\Be   = "0522  
\mathchardef\Bee  = "050F  
\mathchardef\Bz   = "0510  
\mathchardef\Bh   = "0511  
\mathchardef\Bthh = "0512  
\mathchardef\Bth  = "0523  
\mathchardef\Bi   = "0513  
\mathchardef\Bk   = "0514  
\mathchardef\Bl   = "0515  
\mathchardef\Bm   = "0516  
\mathchardef\Bn   = "0517  
\mathchardef\Bx   = "0518  
\mathchardef\Bom  = "0530  
\mathchardef\Bp   = "0519  
\mathchardef\Br   = "0525  
\mathchardef\Bro  = "051A  
\mathchardef\Bs   = "051B  
\mathchardef\Bsi  = "0526  
\mathchardef\Bt   = "051C  
\mathchardef\Bu   = "051D  
\mathchardef\Bf   = "0527  
\mathchardef\Bff  = "051E  
\mathchardef\Bch  = "051F  
\mathchardef\Bps  = "0520  
\mathchardef\Bo   = "0521  
\mathchardef\Bome = "0524  
\mathchardef\BG   = "0500  
\mathchardef\BD   = "0501  
\mathchardef\BTh  = "0502  
\mathchardef\BL   = "0503  
\mathchardef\BX   = "0504  
\mathchardef\BP   = "0505  
\mathchardef\BS   = "0506  
\mathchardef\BU   = "0507  
\mathchardef\BF   = "0508  
\mathchardef\BPs  = "0509  
\mathchardef\BO   = "050A  
\mathchardef\BDpr = "0540  
\mathchardef\Bstl = "053F  
\newcount\rifer \rifer=0 \newcount\trans \trans=0
\def\LAB#1{\advance\rifer by 1 ${}^{\bf t\number\rifer}$}
\def\raf#1{\advance\trans by 1 ${}^{\bf t\number\trans}$}
\begin{document}
\maketitle

\begin{abstract}\0From the ``vibrating string'' and ``Kepler's equation''
 theories to relativistic quantum fields, (divergent)
 series resummations, perturbation theory, KAM theory.\end{abstract} \*

\0{\small {\bf Key words:} \it Vibrating string, Quantum fields,
  Perturbation theory, Divergent series resummation, KAM }%
\footnote[0]{{\small\texttt giovanni.gallavotti@roma1.infn.it}  \small
  extended version of a talk at the ``{\it Il segno di Lagrange nella Matematica
  contemporanea}'', Torino, Politecnico, 10 April 2013.\label{n1}}
\*

\def\SEC{Law od continuity for the vibrating string}
\section{\SEC}\label{sec1}
\iniz 

In 1759 an important open problem was to establish the correctness of
Euler's solution of the wave equation for a string starting from an initial
configuration with a given shape $u_0(x)=\f(x)$ and no initial velocity:

\be\eqalign{
\dpr_t^2 u=&c^2 \dpr^2_x u, \qquad x\in[0,a],\ u(a)=u(b)=0\cr
u(x,t)=& \f(x-ct)+\f(x+ct)
\cr}\Eq{e1.1}\ee
According to D'Alembert's arguments it should have been necessary that
$\f(x)$ be at least a {\it smooth} function (infinitely 
differentiable in present notations or, in D'Alembert's language, subject
to the {\it loi de continuit\'e}) of $x\in R$, {\it periodic with
  period $2a$} and odd around $0$ and $a$. In his notations $\f$ had to
satisfy conditions at $0,a$ allowing its continuation to a $2a$--periodic
function of class $\CC_\infty(R)$, odd around $0$ and $a$ (the arguments
really would only require class $\CC_2(R)$),

Euler claimed that Eq.\equ{e1.1} would be a solution
$u(x,t)=\f(x-ct)+\f(x+ct)$ if $\f(x)$ was smooth in $[0,a]$ and simply
defined outside the interval $[0,a]$ by just continuing it as a periodic
function of period $2a$ odd around $0$ and $a$. He insisted that the
smoothness in $[0,a]$ of $\f(x)$ is sufficient: this means that as
time $t$ becomes $>0$ the form of the string appears (in general) to keep
points with discontinuous curvature: {\it i.e.} points in which the second
derivative is not defined. This led D'Alembert to think that when the
continuation of $\f$ was not smooth the solutions did not make sense.

Neither was able to produce a rigorous argument. Other theories were due to
Taylor who (starting from an initial configuration in which the string, at
rest, is given a shape $u_0(x)=\f(x)$) had proposed that a general solution
of the string motion is a sum $u(x)=\sum_{n} \a_n \sin(\frac{2\p}{2a} n
x)\cos(\frac{2\p}{2a} n c t)$: a view supported by D. Bernoulli (according
to Lagrange). This is criticized by Euler who objects that such expressions
only represent D'Alembert's solutions (which is true {\it if the series
  converges} in class $\CC_2(R)$).

All solutions had been obtained by arguments relying on unproved properties
based on intuition or experience and this is the reason identified by
Lagrange as the source of the controversies. 

Lagrange's idea is that the problem must first be (in modern locution) {\it
  regularized}: this means imagining the string to consist of an indefinite
number $m$ of small particles aligned (in the rest state) and elastically
interacting with the nearest neighbors possibly moving orthogonally to the
rest line and with extremes $y_0=y_m=0$ fixed.

In this way the problem becomes a clearly posed mechanical problem and the
equations of motion are readily derived: the only question is therefore
finding the properties of their solutions and of the limit in which the
mass of the particles tends to zero while their number and the strength of
the elastic force tend to infinity so that a continuum string motion
emerges, \cite[T.I, p.71]{La867}:\footnote[1]{English translation of
  quotations in the endnotes correspondingly labeled {\bf{t}*}}
\* 
\0{\it Il resulte de tout cet expos\'e que l'Analyse que nous avons
  propos\'ee dans le Chapitre pr\'ec\'edent est peut-\^etre, la seule qui
  puisse jeter sur ces mati\`eres obscures une lumi\`ere suffisante \`a
  \'eclaircir les doutes qu'on forme de part et d'autre.}\LAB{c1}
\*

The analysis is today well known: he notices that the problem is (in modern
language) the diagonalization of a $(m-1)\times(m-1)$ tridiagonal symmetric
matrix. The solution is perhaps the first example of the diagonalization
procedure of a large matrix. The result is the representation of the general
motion of the chain with general initial data for positions {\it and
  velocities}, \cite[T.I, p.97]{La867}:

\*
\0{\it .. je ne crois pas qu'on ait jamais donn\'e pour cela une formule
  g\'en\'erale, telle que nous venons de la trouver.}\LAB{c2} see however
\cite{WC987}. 
\*

Lagrange's very remarkable three memories, the first two constitute a
veritable monograph, consist in showing (in full detail and rigor) that the
general motion has, for suitable $\o_h$, the form (translated in modern
language)
\be  y^{(\d)}(\x,t)=\sum_{h=1}^{m-1}
\Big\{
\wt A_h\sqrt{\frac2m}\sin\frac{\p\,h}a\x\,\cdot\cos\o_h t+
\wt B_h\sqrt{\frac2m}\sin\frac{\p\,h}a\x\,\cdot\sin\o_h
t\Big\},\Eq{e1.2}\ee
where, denoting by $\d$ (called $dx$ by Lagrange) the mesh of the discretized
positions so that the number of small masses, located at points $\x=i\d$,
is $\frac{a}\d$ and $ \wt A_h,\wt B_h,\o_h$ are derived from the initial
profiles of positions $Z(\x)$ and velocities $U(\x)$, \cite[T.I.,
  p.163]{La867}: via the expressions

\be \kern-3mm\eqalign{
&\sqrt{\frac2m}\wt A_h= \frac2m \sum_{i=1}^{m-1} \big(\sin\frac{\p h}a\x\big)
Z(\x)\tende{\d\to0}\frac2a\int_0^a Z(x) \big( \sin\frac{\p h}a
x\big)\,dx\cr
&\sqrt{\frac2m}\wt B_h= \frac2{\o_h\,m}
\sum_{i=1}^{m-1}\big(\sin\frac{\p h}a\x\big)
U(\x)\tende{\d\to0}
\frac2{\lis\o_h\,a}\int_0^a U(x) \big( \sin\frac{\p h}a
x\big)\,dx,\cr
&\o_h=c\sqrt{2\frac{1-\cos(\frac{\p h\d}{a})}{\d^2}}\tende{\d\to0}
\lis\o_h=c\frac{\p h}a}\Eq{e1.3}\ee
where $c$ is $\sqrt{\frac\t\m}$, with $\t$ the tension and $\m$ the density
of the string. The formulae Eq.\equ{e1.3} do not require smoothness
assumptions on the data $Z,U$ other than, for instance, continuity as
stated in answer to D'Alembert's critique, \cite[T.I, p.324]{La867}: \*
%
\0{\it Mais je le prie de faire attention que, dans ma solution, la
  d\'etermination de la figure de la corde \`a chaque instant d\'epend
  uniquement des quantit\'es $Z$ et $U$, lesquelles n'entrent point dans
  l'op\'eration dont il s'agit. Je conviens que la formule \`a laquelle
  j'applique la m\'ethode de M. Bernoulli est assujettie \`a la loi de
  continuit\'e; mais il ne me parait pas s'ensuivre que les quantit\'es $Z$
  et $U$, qui constituent le coefficient de cette formule, le soient aussi,
  comme M. d'Alembert le pr\'etend.}\LAB{c3}
\*

Hence, setting $\x=x$, the general solution:
\be \eqalign{ u(x,t)=&\sum_{h=0}^\infty\sin\frac{\p h}a x \,\Big\{ (\frac2a
  \int_0^a Z(x') \sin\frac{\p h}a x'\, dx')\cos\lis\o(h) t\cr &+ (\frac2a
  \int_0^a U(x') \sin\frac{\p h}a x'\,dx') \frac{\sin\lis\o(h)
    t}{\lis\o(h)}\Big\}\cr}\Eq{e1.4}\ee
is found: and it is far more general than the previous solutions because it
permits initial data with $U\ne0$, {\it i.e.} with initial data in which
the string is already in motion. In the case $U=0$ it can be written
$\f(x+ct)+\f(x-ct)$ with $\f$ odd around $0$ and $a$ and $2a$-periodic
aside from the obvious convergence and exchange of sum and limits problems.

Check of Eq.\equ{e1.3} is based on trigonometric properties: basically
on ``Cote's formula'', \cite[T.I, p.75]{La867},
$|a^m-b^m|=\prod_{h=0}^{m-1} (a^2-2ab \cos\frac{2\p}mh+b^2)^{\frac12}$
(derivable from $a^m-b^m= \prod_{h=0}^{m-1} (a e^{i\frac{2\p}m h}-b)$)
which is used instead of the modern $\sum_{h=0}^{m-1} e^{i\frac{2\p}m
  h(p-q)}=m \d_{p,q}$).

The continuum limit Eq.\equ{e1.4} is identified with Euler's result: it is
remarkably found together with interesting considerations and attempted
justifications of resummations of divergent series like, \cite[T.I,
  p.111]{La867},

\be \cos x+\cos 2x+\cos 3x +\ldots=-\frac12\Eq{e1.5}\ee
for $x\ne0$, immediately criticized by D'Alembert, \cite[T.I,
  p.322]{La867}.%
\footnote[2]{\small Lagrange uses this formula to infer that for $t=0$ the
  string shape is described by the function $Z$ and the speed configuration
  by $U$. The argument amounts to proving $\frac1a\sum_{h=1}^\infty
  \sin\frac{\p}ahx\sin\frac{\p}ahy=\d(x-y)$: looking at Lagrange's theory
  it appears, as he correctly exposes in \cite[T.I, p.111]{La867}, that
  what is really used is the truncated version of the above relations and
  the Eq.\equ{e1.5} only plays the role of an intermediate illustrative
  proposition: {\it il ne sera pas hors de propos de d\'emontrer encore la
    m\^eme proposition d'une autre mani\`ere}, \cite[T.I, p.109]{La867}
  (``it will not be out of place to prove again the same proposition by
  another method''). The analysis essentially repeats in different form the
  just proved completeness of the trigonometric sines basis in the finite
  $m$ case; it was derived at a time when even the notion of continuous
  functions (not to mention of distributions) was not formalized, is
  masterful although not formal by our standards; and is used also to infer
  that for $t>0$ the solution coincides with Euler's: for this purpose the
  bold statement that $\sin m(\frac\p{a}({x}\pm c t))=0$ if $m=\infty$ is
  made and better justified in the answer to D'Alembert quoted above after
  Eq.\equ{e1.3}, \cite[T.I, p.324]{La867}, although D'Alembert's questions
  are not really answered also because of the needed further exchange of
  limits intervening to replace $\o_h\, t$ with $h\,c\,
  t\frac\p{a}$.\label{n2}}

To D'Alembert's objections Lagrange later gave also new arguments,
apparently ``far less'' convincing: \* 
\0{\it Or je demande si, toutes les fois que dans une formule alg\'ebrique
  il se trouvera par exemple une s\'erie g\'eom\'etrique infinie, telle que
  $1 +x + x^2+x^3+x^4+\ldots$, on ne sera pas en droit d'y substituer
  $\frac1{1-x}$ quoique cette quantit\'e ne soit r\'eellement \'egale \`a
  la somme de la s\'erie propos\'ee qu'en supposant le dernier terme
  $x^\infty$ nul. Il me semble qu'on ne saurait contester l'exactitude
  d'une telle substitution sans renverser les principes les plus communs de
  l'analyse}\LAB{c4}
\*\0\cite[T.I, p.323]{La867}, where the comment refers to the case $x\ne1$
and illustrates the conceptions on divergent series common already at the
time.\footnote[3]{The case $x=1$ was discussed in the original paper,
\cite[T.I, p.111]{La867}: \\
\0{\it Mais, dira-t-on, comment peut-il se faire que la somme de la suite
  infinie $\cos x+ \cos2x+ \cos3x+...$ soit toujours \'egale \`a $-\frac12$
  puisque, dans le cas de $x = 0$, elle devient n\'ecessairement \'egale
  \`a une suite d'autant d'unit\'es? Je r\'eponds que ...}
\label{n3}}\ \LAB{c5}
And, about a further similar criticism by D'Alembert, applying to the
alternating series obtained setting $x=\frac\p4$: \*
\0{\it Je r\'eponds qu'avec un pareil raisonnement on
  soutiendrait aussi que $\frac1{1+x}$ n'est point l'expression
  g\'en\'erale de la somme de la suite infinie $1-x+x^2-x^3+\ldots$ parce
  que, en faisant $x=1$, on a $1-1+1-1+\ldots$ ce qui est, ou $0$, ou $1$,
  selon que le nombre des termes qu'on prend est pair ou impair, tandis que
  la valeur de $\frac1{1+x}$ est $\frac12$. Or je ne crois pas qu'aucun
  G\'eom\`etre voul\^ut admettre cette conclusion}\LAB{c6} \* \0\cite[T.I,
  p.323]{La867}.

The analysis might appear not rigorous in today sense: not really because
of the statement Eq.\equ{e1.5}; nor because of $\sin (\frac\p{a} m(x\pm
ct))=0$ as consequence of $m=\infty$, \cite[T.I, p.102]{La867}, see
footnote{${}^{\ref{n2}}$}: which may make the eyebrows frown, as of course
D'Alembert's did, and for which Lagrange almost ``apologized'': \*

\0``{\it Je conviens que je ne me suis pas exprim\'e assez exactment ..
}\LAB{c7}'',

\*\0\cite[T.I, p.322]{La867}, while not making it more plausible
(it seems to be an early version of what is today the ``Lebesgue theorem''
for Fourier series).

Also several other objections by D'Alembert and D. Bernoulli do not appear
to have been answered very convincingly, by the present standards, in the
memory in defense of the theory in \cite[T.I, p.319-332]{La867}. The short
note is still of great interest as it shows that Lagrange struggles and gets
very close to the modern notion of ``weak solution'' of a PDE, see also
\cite[T.I, p.177]{La867}, and to a formalization of a theory of resummation
of divergent series. It is unlikely that D'Alembert was ever convinced by 
Euler or Lagrange: but Lagrange was strongly supported by
Euler,\cite[T. XIV, p.111]{La867}, writing:
\*
\0{\it je vous avoue qu'elles ne me paraissent pas assez fortes pour renverser
  votre solution. Ce grand g\'enie me semble un peu trop enclin \`a
  d\'etruire tout ce qui n'est pas construit par lui-même.  \\ ...
  \\ Apr\`es cette remarque, je vous accorde ais\'ement, monsieur, que
  pourque le mouvement de la corde soit conforme \`a la loi de
  continuit\'e, il faut que, dans la figure initiale, les $\frac{d^2
    y}{dx^2},\frac{d^4 y}{dx^4}, \frac{d^6 y}{dx^6}$, soient \'egales \`a
  $0$ aux deux extr\'emit\'es; mais, quoique ces conditions n'aient pas
  lieu, je crois pouvoir soutenir que notre solution donnera n\'eanmoins le
  v\'eritable mouvement de la corde; car ...}\LAB{c8}
\*

In conclusion today a mathematician or a physicist will probably consider
it not completely rigorous only because although it applies to twice
$\CC_2([0,a])$ initial data $Z,U$, for the positions and speeds of the
string elements, nevertheless a proof of the exchanges of limits (and their
existence) needed to ``pass to the continuum limit'' is not even
mentioned. As it would be expected since the condition is not met at $t>0$
as the discontinuity in the second derivatives (if present at $t=0$)
migrates from the boundary to points moving with the wave in the interior
of $[0,a]$ (which is probably what really worried D'Alembert).

The simple (sufficient) condition, implying the existence of the continuum
limit, that initial data vanishing at the extremes should also have two
continuous derivatives became clear only later when Fourier established the
theory of trigonometric series.\footnote[4]{Thus obtaining a rigorous
  extension beyond D'Alambert's ``continuity condition'' which required two
  continuous derivatives everywhere, vanishing second derivative at the
  extremes and a string initially deformed but at rest.\label{n4}}

Its need did not occur to Lagrange, in his 23-d year of age, nor it did
occur to Euler himself: the easy proof is now in all textbooks, {\it e.g.}
\cite[Ch.4.5]{Ga008} for the D'Alembert's case, and it is rightly
considered that, {\it de facto}, Lagrange gave a key argument that later
led to Fourier's solution of the controversies on the proof of completeness
of the basis $\sin \frac{\p}{a}nx$ for the functions on $[0,a]$ vanishing
at $0$ and $a$, {\it i.e} proved, for instance, the convergence of their 
Fourier's series when twice continuously differentiable. For a detailed
discussion of the vibrating string controversy with  the points of view of
D'Alembert, Euler, Bernoulli see \cite{WC987}.

Having determined completely the motion of a discrete chain with arbitrary
initial conditions was enough to justify the ``extension'' to the continuum
limit, non rigorous by standards that were just beinning to emerge, as
D'Alembert's objections witness.

The theory of the vibrating strings, in Lagrange, was motivated and became
part of a long and detailed study of the propagation of sound in two
exhaustive Memoires (followed by a short one to answer criticism): in which
sound propagation, \Ie the wave equation, is analyzed in the one
dimensional case and for three dimensional spherical waves (whose equation
is taken from Euler and which he reduces to the one dimensional theory via
the remarkable change of variables $u(x)=x^{-2}{\int^x z(x)x dx}$).

In the second Memory the law of continuity problem is also examined from a
new viewpoint: namely to study the motion of the {\it internal points} of a
string with extremes fixed. It is proved that the points move following the
solution found with the discretization method (in spite of the travel of
the continuity break through $[0,a]$). This is interesting also because the
method presented will be recognized to be a precursor of the modern notion
of {\it weak solution} of a PDE, \cite[T.I, p.177]{La867}: \*

\0{\it Les transformations dont je fais usage dans cette occasion sont celles
qu'on appelle int\'egrations par parties, et qui se d\'emontrent 
ordinairement par les principes du calcul diff\'erentiel; mais il n'est pas
difficile de voir qu'elles ont leur fondement dans le calcul g\'en\'eral des
sommes et des diff\'erences; d'o\`u il suit qu'on n'a point \`a craindre
d'introduire par l\`a dans notre calcul aucune loi de continuit\'e entre les
diff\'erentes valeurs de $z$.}\LAB{c9}
\*

The vibrating string theory and the continuum as a limit of a
microscopically discrete reality was developed by Lagrange at a time when
the atomistic conceptions were being established. The approach he adopted
is a key legacy and the basis of methods currently employed in the most
diverse fields, see for instance \cite{Pr009}: a further example will be
discussed in the next section.
 

\def\SEC{QFT: quantum elastic string}
\section{\SEC}
\label{sec2}\iniz

The key idea in the vibrating string has been that it is a continuous
system which should be regarded as, and behaves as, a limit case of a
system of infinitely many adjacent particles whose motion should be described
via the ordinary equations {\it without} requiring new principles.

Interestingly this problem has reappeared essentially for the same reasons
in recent times. The theory of elementary particles requires at the same
time quantum mechanics and (at least) special relativity: it became soon
clear that it could be appropriately formulated as a theory of quantized
fields which met immediately impressive successes in the description of
electromagnetic interactions of photons and electrons and of weak
interactions. Particles were naturally represented by particular states of
a field which could describe waves as well, \cite[Sec.I]{Ga985b}.

The simplest example is a ``scalar field'', $\f(x)$, in space time
dimension $2$ corresponding classically to the Lagrangian
\be\eqalign{
{\cal L}=&\frac\m2\int_\a^\b \Big(\dot\f(x)^2- c^2(\frac{d\f}{d
    x}(x))^2
-(\frac{m_0 c^2}{\hbar})^2\f(x)^2\,-\, I(\f(x))\Big)\,dx
\cr}\Eq{e2.1}\ee
where $I(\f)$ is some function of $\f$. If $I=0$ this is a vibrating string
with density $\m$, tension $\t=\m c^2$ and an elastic pinning force $\m
(\frac{m_0 c^2}\hbar)^2$.

\eqfig{300}{65}{
\ins{-15}{0}{$O$}
\ins{260}{0}{$L$}
}{Fig1}{Fig.1}

\kern4mm
\0{\small Figure 4.1: chain of oscillators elastically bound by nearest
  neighbors and to centers aligned on an axis orthogonal to their vibrations
and Dirichlet's boundary condition.}
\*

The nonlinearity of the associated wave equation produces the result
that when two or more wave packets collide they emerge out of the
collision quite modified and do not just go through each other as in
the case of the linear string, so that their interaction is
nontrivial. 

The model is naturally generalized to space-time dimension $D+1$ if $x$ is
imagined a point $\V x$ of a cubic lattice in a $D$-dimensional cube and
$\f(\V x)$ describes the deformation of an elastic film ($D=2$) or body
($D=3$).

Naively the quantum states will be, by the ``natural extension of the usual
quantization rules'', \Ie functions $F(\f)$ of the profile $\f$ describing the
configurational shape of the elastic deformations. The Hamiltonian operator
acts on the wave function $F$ as
\be\eqalign{
({\cal H} F)(\f)=&
\int_{0}^L
\left(-\frac{\hbar^2}{2\m}
\frac{\d^2 F}{\d\,\f(\V x)^2}(\f)+
\frac\m2\Big(
c^2
(\frac{\dpr\f}{\dpr\V x}(\V x))^2+\right.\cr
&\left.+(\frac{m_0 c^2}{\hbar})^2 
\f(\V x)^2+ I(\f(\V x))\Big)\, F(\f)\right)
\,d^D\V x
\cr}\Eq{e2.2}\ee
where $\frac\d{\d\f(\V x)}$ is the functional derivative operator (a notion
also due to Lagrange and to his calculus of variations) 
and it should be defined in the space $L_2(``d\f'')$, where the scalar
product ought to be $(F,G)=\int \overline{F(\f)}\,G(\f)\,''d\f''$ and
$''d\f''=\prod_{\V x\in[0,L]} d\f(\V x)$.

Even though by now the mathematical meaning that one should try to attach
to expressions like the above, as ``infinite dimensional elliptic
operators'' and ``functional integrals'', is quite well understood,
particularly when $I\equiv0$, formulae like the above are still quite
shocking for conservative mathematicians, even more so because they turn
out to be very useful and deep.

One possible way to give meaning to \equ{e2.2} is to go back to first
principles and recall the classical interpretation of the vibrating string
as a system of finitely many oscillators, following Lagrange's brilliant
theory of the discretized wave equation and of the related Fourier series
summarized in Sec.\ref{sec1}.

Suppose, for simplicity, that the string has periodic boundary conditions
instead of the Dirichlet's conditions studied by Lagrange; replace it with
a lattice $ Z_\d$ with mesh $\d>0$ and such that $L/\d$ is an integer. In
every point $n \d$ of $ Z_\d$
\\(1) locate an oscillator with mass $\m\,\d$,
described by a coordinate $\f_{n\d}$ giving the elongation of the
oscillator over its equilibrium position,
\\(2) subject to elastic pinning
force with potential energy $\frac12 \m \d\,(\frac{m_0 c^2}{\hbar})^2
\f^2_{n\d}$,
\\
(3) to a nonlinear pinning force with potential energy $\frac12\m
\d I(\f_{n\e})$
\\
(4) finally to a linear elastic tension, coupling nearest
neighbors at positions $n\d,(n+1)\d$, with potential energy $\frac12\m
\d^{-1}c^2 (\f_{n\d}-\f_{(n+1)\d})^2$.

Therefore the Lagrangian of the classical system, in the more general case
of space dimension $D\ge1$ ($D=1$ is the vibrating string, $D=2$ is the
vibrating film, {\it etc}, \Ie space-time dimension $d=D+1$), is
\be\kern-3mm\eqalign{
{\cal L}=&\frac\m2 \d^D\sum_{n \d\in \L_0}\Big(
\dot\f_{n\d}^2-c^2 \sum_{j=1}^D \frac{(\f_{n \d+ e_j \d}-
\f_{n\d})^2}{\d^2}
-(\frac{m_0 c^2}{\hbar})^2\f^2_{n\d}- I(\f_{n\d})\Big)
\cr}\kern-3mm
\Eq{e2.3}\ee
where $e_j$ is a unit vector oriented as the directions of the
lattice; if $n \d+e_j \d$ is not in $\L_0$ but $n \d$ is in
$\L_0$ then the $j$-th coordinate equals $L$ and $n
\d+e_j \d$ has to be interpreted as the point whose $j$-th coordinate
is replaced by $1$; {\it i.e.} is interpreted with periodic
boundary conditions with coordinates identified modulo $L$.

It should be remarked that for $I=0$ and $m_0=0$ and Dirichlet boundary
conditions Eq.\equ{e2.3} is, for $D=1$, the Lagrangian of the discretozed
vibrating string introduced by Lagrange in his theory of sound.

Of course there is no conceptual problem in quantizing the system: it will
correspond to the familiarly elliptic operator on $L_2(\prod_{n \e} d\f_{n
  \e})=L_2(R^{\frac{L}\d})$:
\be\eqalign{
\HH_{\d}=&
-\frac{\hbar^2}{2\m \d}\sum_{n\d\in\L_o}\frac{\dpr^2}{\dpr \f_{n
\d}^2}+\frac{\m \d}{2}
\sum_{n \d\in\L_o}\cr
&\Big( c^2\frac{(\f_{n \d+e
    \d}-\f_{n \d})^2}{\d^2}+
\big(\frac{m_0 c^2}{\hbar}\big)^2 \f_{n \d}^2+ I(\f_{n \d})\Big)
\cr}\Eq{e2.4}\ee
with $\CC^\infty_0(R^{\frac{L}\d})$ as domain (of essential
self-adjoint\-ness) provided $I(\f)$ is assumed bounded below, as it should
always be: 
\eg in the case of the so called {\it $\l \f^4$ theory} with
$I(\f)=\int_0^L (\l \f(x)^4+\m\f(x)^2+\n)dx$ with $\l>0$.

At this point it could be claimed, with Lagrange \cite[T.I, p.55]{La867},
\*

\0{\it Ces \'equations, comme il est ais\'e de le voir, sont en m\^eme
  nombre que les particules dont on cherche les mouvements; c'est pourqoi,
  le probl\`eme \'etant d\'ej\`a absolument determin\'e par leur moyen, on
  est oblig\'e de s'en tenir l\`a, de sorte que toute condition
  \'etrang\`ere ne peut pas manquer de rendre la solution insuffisante et
  m\^eme fautive.}\LAB{c10}
\*

This can be appreciated by recalling that in developing the theory, {\it
  even in the simplest case of the $\l\f^4$ field, just mentioned},
difficulties, which generated many discussions, arise.

In particular the attempts to study the properties of the operator $\cal H$
via expansions in $\l,\m,\n$ immediately lead to nonsensical results
(infinities or indeterminate expressions). At the beginning of QFT the
results were corrected by adding ``{\it counterterms}'' amounting, in the
case of the string {\it i.e.} of space-time of dimension $2$, to make
$\m,\n$ infinite in the continuoum limit (\Ie as $\d\to0$), suitably fast. 

The subtraction prescriptions, known as ``{\it renormalization}'' were not
really arbitrary: the remark was that {\it all results} were given by many
integrals which were divergent but in which a divergent part could be
naturally isolated by bounding first the integration domains and by
determining $\m,\n$ as functions of $\d$ and of the parameters defining the
domains boundaries so that the infinities disappear when the boundaries of
the domains of integration are removed.

{\it Reassuringly} the choice of the divergent ``counterterms'' $\m,\n$
turned out to be essentially independent of the particular result that was
being computed.

Even though the {\it renormalization} procedure was unambiguous, at least
it was claimed to be such in the physically more interesting model of
quantum electrodynamics as Feynman states in the abstract of \cite{Fe949}:
\*
\0{\it A complete unambiguous and presumably consistent, method is therefore
available for the calculation of all processes involving electrons and
photons}
\*
\0it was not clear (at least not to many) that
some new rule, \Ie a new physical assumption, was not implicitly introduced
in the process. For instance Dyson:
\*
\0{\it Finally, it must be said that the proof of the finiteness and
  unambiguity of $U(\infty)$ given in this paper makes no pretence of being
  complete and rigorous. It is most desirable that these general arguments
  should as soon as possible be supplemented by an explicit calculation of
  at least one fourth-order radiative effect, to make sure that no
  unforeseen difficulties arise in that order}, \cite[p.1754]{Dy949b},
\*
\0or similarly, later, Heitler.\footnote[5]{\it 
On the one hand we can state that the present theory cannot be final. We
have found a number of divergent quantities, although all of them are
unobservable in principle. These are ...
\\
Furthermore, we found in some cases that even observable effects are
described by ambiguous mathematical expressions...
\\
The ambiguities can always be settled by applying a certain amount of
'wishful mathematics', namely by using additional conditions for the
evaluation of such ambiguous integrals...
\\
On the other hand, these difficulties do not prevent us from giving a
theoretical answer to every legitimate question concerning observable
effects. These answers are, whenever they can be tested, always in
excellent agreement with the facts..., \cite[p.354]{He953}.}

This is analogous to the controversy on the vibrating
string solved by Lagrange with his theory of sound via
\\(1) discretization of
the string,
\\(2) solution of its motion and
\\
(3) removal of the regularization, \Ie
taking the particles mass to $0$, the tension to $\infty$ at proper rates
as $\d\to0$ so that all results remained well defined and converging to
limits.

In the 1960's reducibility of renormalization theory to the rigorous study
of the properties of the operator in Eq.\equ{e2.4} was started and at least
in the above cases of the quantum string and of the quantum film (\Ie the
cases of Eq.\equ{e2.4} with space-time dimension $2$ and $3$, respectively)
it was fully understood in a few (very few) cases via the work of Nelson,
Glimm-Jaffe, Wilson who showed clearly that no infinities really arise if
the problem is correctly studied, \cite{Ga985b}: \Ie first taking seriously
the discretized Hamiltonian, then computing physically relevant quantities
and finally passing to the continuum limit, just as Lagrange did in his
theory of sound.

This has been a major success of Physics and Analysis. The problem remains
open in the case of $d=D+1=4$ space-time dimensions and the reason is precisely
that the nature of the regularization becomes essential in higher
dimension: it seems that the naive discretization on a square lattice
described here implies assumptions on the Physics at very small scale so
that it might be impossible to perform the continuum limit $\d\to0$ unless
the regularization of the formal functional derivatives in the Hamiltonian
is chosen conveniently. It is conjectured that a naive discretization like
the above Eq.\equ{e2.4} cannot lead to a nontrivial result if $d=4$ and
$I(\f)$ is a fourth order polynomial.

Therefore a challenge remains to explain why the renormalized quantum
electrodynamics in $4$-space time dimensions, only defined as a formal
power expansion in the elementary electric charge, gives amazingly precise
results perfectly agreeing with experiments, although the known
regularizations are all conjectured to yield trivial result if studied
under full mathematical rigor. In recent times the regularization choice
appears to have been related to very new conceptions of the nature at very
small space-time scale (QCD and string theory): even in this respect
Lagrange's method of attacking problems avoiding the introduction of new
principles or of understanding their necessity is still fertile.

\def\SEC{Perturbations in Field Theory}
\section{\SEC}
\label{sec3}\iniz

The problems of QFT are often studied by perturbation theory and a key role
is played by {\Ie scale invariant theories}. For instance consider the
elliptic operator $\HH_\d$ in Eq.\equ{e2.4} for a quantum string, written
as $H^0_\d+I_\d$ separating the term with
$I(\f_{n\d})=\l\f_{n\d}^4+\m\f_{n\d}^2+\n$, \Ie with the interaction.

Perturbation theory as developed by Lagrange might seem at first
sight quite far from its use in QFT: yet it is quite close as it will be
discussed in this section after giving some details on the form in which it
arises in renormalization theory and how it appears as an implicit
functions problem.

Restricting the discussion to the scalar $\f^4$-systems, Eq.\equ{e2.4}, the
breakthrough has been Wilson's theory which in particular shows that the
values of physical observables can be can be constructed, for the $\f^4$
system in dimensions $d=2,3$, as power series in a sequence of parameters $
\L_k\defi(\l_k,\m_k,\n_k)$, $k=0,$ $1,2,\ldots$, called {\it running
  coupling constants}, which are related by

\be
\L_k= M \L_{k+1}+ B(\{\L_r\}_{r=k+1}^{\infty}),\qquad k=0,1,\ldots\Eq{e3.1}\ee
where $M$ is a diagonal matrix with elements $m_1,m_2,m_3$; $B,M$ define,
via Eq.\equ{e3.1} two operators $\BB,\MM$ on the sequences
$\{\L_r\}_{r=0}^{\infty}$.

The physical meaning of the running constants is that they control the
physical phenomena that occur on length scale $2^{-k} \ell_0$, where
$\ell_0$ is a natural length scale associated with the system, \eg
$\ell_0=\frac{\hbar}{m_0 c^2}$.

The values of observable quantities are power series in the running
constants which, order by order, are {\it well defined and no infinities
  appear provided} the sequence $\{\L_r\}_{r=0}^\infty$ consists of
uniformly bounded elements.

Hence the physical observables values can be expressed formally in terms of
a well defined power series in few parameters, the $\L_k$'s, which are in
turn functions of three independent parameters (the $\L_0$, for instance,
often called the ``physical'' or ``dressed couplings'')
whose values define the theory. 

However the values of physical observables are {\it very singular
  functions} of the $\L_r$ and their expansion in powers of the physical
parameters (\Ie $\L_0$, for instance) would be meaningless. In other words
the infinities appearing in the heuristic theory are due to singularities
of the $\L_r$'s as functions of a single one among them: in other words
they are due to overexpanding the solution.

This leaves however open the problems:

\0(1) existence of a bounded sequence of ``running couplings'' $\L_k$

\0(2) convergence of the series for the observables values

The second question has a negative answer because it is immediate that
$\l_k<0$ cannot be expected to be allowed: however in the $\f^4$-systems
they can be shown to be asymptotic series in space-time dimensions $2,3$.

The first question has an answer in space-time dimensions $2,3$ (ultimately
due to $m_1<1$ and $m_2,m_3>1$) while in $4$ dimensions space-time one more
expansion parameter $\a_k$ is needed and $m_1,m_4=1$ but it seems
impossible to have a bounded sequence with $\l_k>0$ (the {\it triviality
  conjecture} proposes actual impossibility, but it is a delicate and open
problem).

A further relation with Lagrange's work comes from Eq.\equ{e3.1}
considered as an equation for $\L\defi \{\L_k\}_{k=0}^\infty$ of the form
$\L-\MM(\L)-\BB(\L)=0$. In the development of perturbation
theory (and in cases more general than the present) the request of a
bounded solution of the latter implicit function problem arises in a form
that has strong similarity with the solution algorithm proposed in
Lagrange's theory of ``litteral equations'' (see below).

In \cite[T.III, p.25]{La867} the following formula is derived
\be\a=x-\f(x),\quad\otto\quad
\ps(x)=\ps(\a)+ \sum_{k=1}^\infty \frac1{k!}
\dpr^{k-1}_\a(\f(\a)^k\dpr_\a\ps(\a))
\Eq{e3.2}\ee
{\it for all} $\ps$. In particular for
$\ps(\a)\equiv\a$ 
\be x(\a)=\a+\sum_{k=1}^\infty \frac1{k!}
\dpr^{k-1}_\a(\f(\a)^k),\qquad x(0)=\sum_{k=1}^\infty \frac1{k!}
\dpr^{k-1}_\a(\f(\a)^k)\Big|_{\a=0}\Eq{e3.3}\ee
are expressions for the inverse function and, respectively, for a solution
of $x=\f(x)$ (for the formula to work the series convergence
is necessary).

In general $x(0)$ in Eq.\equ{e3.3} does not always select among the roots
the closest to $\a$ as pointed out in \cite{Ch846}, where a
related statement by Lagrange is criticized and a discussion
is given of the properties of the root obtained by replacing $\f$ by $t \f$
and studying which among the roots is associated with the sum of the series
if the latter converges up to $t=1$, theorem 3, p.18.

The formulae can be generalized to $\a,x$ in $R^n,\, n\ge1$ and are used in
the derivation and on the detailed analysis of Eq.\equ{e3.1} via the remark
that the fixed point equation of Lagrange is equivalent to the following
{\it tree expansion} of the $k-th$ term in Eq.\equ{e3.2} (when $\ps(x)=x$):
\be \frac1{k!} \dpr^{k-1}_\a(\f(\a)^k)=\sum_\th {\rm Val}(\th)\Eq{e3.4}\ee
where $\th$ is a tree graph, \Ie
it is a ``decorated'' tree, see Fig.2, with 
\\
(1) $k$ branches $\l$ of
equal length oriented towards the ``root'' $r$, \\ 
(2) each node as well as
the root (not considered a node) carries a label $j_v\in\{1,\ldots,n\}$
\\
(3) at each node $v$ enter $k_v$ branches $\l_1\equiv v
v_1,\ldots,\l_{k_v}\equiv v v_{k_v}$ where $v_1,\ldots,v_{k_v}$ are the
$k_v$ nodes preceding $v$ (it is $\sum_{v<r}k_v=k$).  and \\ (4) the node
$v$ symbolizes the tensor $\dpr^{k_v}_{x_{j_{v_1}},x_{j_{v_2}},\ldots
  x_{j_{v_{k_v}}} }\f_{j_v}(x)$

\eqfig{300}{145}{
\ins{-5}{71}{$r$}
\ins{14}{71}{$j$}
\ins{24}{91}{$\l_{0}$}
\ins{50}{71}{$v_0$}
\ins{127}{100.}{$v_1$}
\ins{92.}{41.6}{$v_2$}
\ins{158.}{83.}{$v_3$}
\ins{191.7}{133.3}{$v_5$}
\ins{191.7}{100.}{$v_6$}
\ins{191.7}{71.}{$v_7$}
\ins{191.7}{-6.3}{$v_{11}$}
\ins{141.7}{20}{$v_{12}$}
\ins{191.7}{18.6}{$v_{10}$}
\ins{166.6}{42.}{$v_4$}
\ins{191.7}{54.2}{$v_8$}
\ins{191.7}{37.5}{$v_9$}
}{Fig2}{Fig.2}

\kern.7cm
\0Fig.2:\ {\small A decorated tree. Labels $j_{v}, v<r$
not marked and intended as
contracted.}
\*
\0(5) Two such decorated trees are considered equivalent if 
superposable by pivoting the branches around the nodes without
permitting overlapping.
\\
\0(6) The {\it value}, ${\rm Val}\,(\th)$ is defined as

\be {\rm Val}\,(\th)=\prod_{v\in\th} \Big(\frac1{k_v!}
\dpr^{k_v}_{j_{v_1},\ldots,j_{v_{k_v}}}\f_{j_v}(x)\Big)\Eq{e3.5} \ee
 where sum over node labels, {\it except the root label $j=j_{r}$},
is understood.
\*

It can be said that also the determination and theory of the functions $B$
follows a path that can be traced to Lagrange's theory of ``litteral
equations'': the formula is also a graphical reformulation of a
combinatorial version of Eq.\equ{e3.3}, \cite{Fa856,Mo013}, and can be used
to derive several combinatorial formulae (like Cayley's count of trees,
\cite[(5.11)]{Ga994}, \cite[p.324]{GG005e}, \cite{Mo013} and as recently
pointed out more, \cite[p.4]{Mo013}).

The analysis in the $\f^4$-Field Theory context of the just described
method of getting to the running couplings expansion can be found in
\cite{Ga985b}: the analogy seems manifest and it would be interesting to
find a closer relation.  

Lagrange made use of Eq.\equ{e3.4} in his works on Celestial Mechanics,
relying on his study of Kepler's equation giving eccentric anomaly $\x$ in
terms of average anomaly $\ell$:

\be \ell=\x+e\sin\x,\qquad \x=\ell+\sum_{k=1}^\infty
\frac{e^k}{k!}\dpr^{k-1}_\ell(\sin^k\ell)
\Eq{e3.6}\ee
The coefficients can be readily evaluated if $\sin^k\x$ is expressed in
terms of angles multiples of $\ell$; and also the true anomaly $u$ can be
computed in powers of the eccentricity via Eq.\equ{e3.2} with $\f(\ell)=e
\sin\ell$ and $\ps(\ell)=\frac{\sqrt{1-e^2}}{1+e\cos\ell}$, because $\frac{
  du}{d\x}=\frac{(1-e^2)^{\frac12}}{1+e\cos\x}$, \cite[T.III,
  p.114]{La867}.

Several applications of perturbation theory have been considered by
Lagrange often leading, as a byproduct, to key discoveries like the
determination of the five Lagrangian points, \cite[T.VI, p.280]{La867},
found while attempting to find approximate solutions to the three body
problem, or the many works on the secular variations of the planetary
nodes, the determination of the orbit of comets, the Moon librations.

\def\SEC{Overview on  Mechanics}
\section{\SEC}
\iniz\label{sec4}\iniz

Lagrange's contribution to the formulation and application of the
principles of Mechanics is so well known and modern that it is continuously
used today. The 1788 {\it M\'ecanique Analytique}, whose second
edition is in \cite[T.XI]{La867} and \cite[T.XII]{La867}, is entirely based
on a consistent reduction to first principles of every problem considered.

This means constantly returning to applying the least action principle in
the form of the combination the principle of virtual works with the
principle of D'Alembert, \cite[p.51]{Da743}, which says: 
\*
\0{\it D\'ecompos\'es les Mouvemens $a,b,c,\&tc$ imprim\'es \`a chaque
  Corps, chacun en deux autres $\a,\a';\b,\b';\g,\g'\&tc$ qui soient tels,
  que si l'on n'e\^ut imprim\'e aux Corps que les Mouvemens $\a,\b,\g,\&tc$
  ils eussent p\^u conserver ces Mouvemens sans se nuire r\'eciproquement;
  et que si on ne leur e\^ut imprim\'e que les Mouvemens $\a',\b',\g',\&tc$
  le syst\`eme fut demeur\'e en repos; il est clair que $\a,\b,\g,\&tc$
  seront les Mouvemens que ces Corps prendront en vertu de leur action. Ce
  Q.F. trouver.}\LAB{c11}
\footnote[6]{The last words, ``What was to be found'', are there because the
  principle is ``derived'' as an answer to a problem posed in the
  previous page. Here $a,b,c,\&tc$ are the impressed forces $\V f_i$, and
  $\a,\b,\g,\&tc$ are $m_i \V a_i$ while $\a',\b',\g',\&tc$ are the
  constraint reactions $-\V R_i$: \Ie $\V f_i=m_i\V a_i-\V R_i$; in modern
  form, \cite[\S3.18,Vol.2]{LA927}, the $\V f_i-m_i\V a_i$ may be called
  the ``lost forces'' and the principle is {\it During the motion of a
    system of point masses, however constrained and subject to forces, at
    each instant the lost forces are equilibrated by virtue of the
    constraints}. } \*

Lagrange's addition is clearly explicitly stated in the second work on the
Moon librations where he says, about his method: \*

\0 {\it... Elle n'est autre chose que le principe de Dynamique de
  M. d'Alembert, r\'eduit en formule au moyen du principe de l'\'equilibre
  appel\'e commun\'ement loi des vitesses virtuelles. Mais la combinaison
  de ces deux principes est un pas qui n'avait pas \'et\'e fait, et c'est
  peut-\^etre le seul degr\'e de perfection qui, apr\`es la d\'ecouverte de
  M. d'Alembert, manquait encore \`a la Th\'eorie de la Dynamique.}\LAB{c12},
\cite[T.V, p.11]{La867}.\footnote[7]{\Ie $\sum_i (\V f_i-m_i\V a_i)\cdot\d\V
  x_i=0$.} \*

The equivalent action principle is used in the form \*

\0{\it Dans le mouvement d'un syst\`eme quelconque de corps anim\'es par
  des forces mutuelles d'attraction, ou tendantes \`a des centres fixes, et
  proportionnelles \`a des fonctions quelconques des distances, les courbes
  d\'ecrites par les diff\'erents corps, et leurs vitesses, sont
  n\'ecessairement telles que la somme des produits de chaque masse par
  l'int\'egrale de la vitesse multipli\'ee par l'\'el\`ement de la courbe
  est un maximum ou un minimum, pourvu que l'on regarde les premiers et les
  derniers points de chaque courbe comme donn\'es, en sorte que les
  variations des coordonn\'ees r\'epondantes \`a ces points soient nulles.
\\
C'est le th\'eor\`eme dont nous avons parl\'e \`a la fin de la premi\`ere
Section, sous le nom de Principe de la moindre action}\LAB{c13}, 
\cite[T.XI, p. 318]{La867}.  \*

He arrives at the {\it M\'ecanique Analytique}, which ``brought
Rational Mechanics to a state of perfection'', \cite[p.221]{Li967}, as the
completion of a long series of remarkable applications. Namely

\0(1) the
theory of sound (\Ie the vibrating string) and the theory of the librations
of the Moon, 1763, \cite[T.VI, p.8]{La867} (revisited in \cite[T.V, p.5]{La867},
1781), which is perhaps the first example of development and systematic
application of pertubation theory based on analytical mechanics, \Ie on the
action principle and the Euler-Lagrange equations.

\0(2) the theory of rigid body, which he undertakes curiously insisting
in avoiding starting from the proper axes of rotation, \cite[T.III,
  p.579]{La867}:
\*

\0{\it ... ce qui exige la r\'esolution d'une \'equation
  cubique. Cependant, \`a consid\'erer le Probl\`eme en lui-même, il semble
  qu'on devrait pouvoir le r\'esoudre directement et ind\'ependamment des
  propri\'et\'es des axes de rotation, propri\'et\'es dont la
  d\'emonstration est assez difficile, et qui devraient d'ailleurs \^etre
  plut\^ot des cons\'equences de la solution m\^eme que les fondements de
  cette solution}\LAB{c14}, \*

\0he rederives the theory in an original way, certainly very involved for
today eyes (used to Euler's equations as beginning point
\cite[p. 402]{Eu853}, \cite{FGG013}), obtaining, as a byproduct, the
theorem about the reality of the eigenvalues of a $3\times3$ symmetric
matrix, \cite[T.III, p. 605]{La867}, and (later, in the {\it M\'canique}) the
motion of the ``Lagrange's top'', \cite[T.XII, p. 253]{La867} \*

\0{\it Ce cas est celui o\`u l'axe des coordonn\'ees $c$, c'est-à-dire la
  droite qui passe par le point de suspension et par le centre de
  gravit\'e, est un axe naturel de rotation, et o\`u les moments d'inertie
  autour des deux autres axes sont \'egaux (art. 32), ce qui a lieu en
  g\'en\'eral dans tous les solides de r\'evolution, lorsque le point fixe
  est pris dans l'axe de r\'evolution.  La solution de ce cas est facile,
  d'apr\`es les trois int\'egrales qu'on vient de trouver}\LAB{c15}, \*

\0a remarkable integrability property that Poisson rediscovered (in fact he
does not quote Lagrange, see the footnote by J. Bertrand at the {\it
  loc.cit.}).

\0(3) the theory of the secular variations of the planetary elements and
several other celestial questions (the fifth volume of the collected paper
is entirely dedicated to celestial mechanics, \cite[T.V]{La867}) opening
the way to Laplace's celestial mechanics and slightly preceding the
completion and publication of the {\it M\'ecanique}. This was the first
time analytical mechanics was used to attempt {\it long time} orbital
predictions.

\0(4) the contributions are not only directed towards the principles but
concentrate on concrete problems like the integration by separation
of variables, \Ie by quadratures, of several dynamical systems: \eg the two
centers of gravitational attraction in the three dimensional case,
\cite[T.II, p.67]{La867} or the mentioned symmetric top.

\0(5) The {\it M\'ecanique Analytique} also reflects the attitude of
Lagrange regarding matter as constituted by particles: \Ie his consistently
kept atomistic view, see for instance \cite[T.XI, p.189]{La867}: \*

\0{\it Quoique nous ignorions la constitution int\'erieure des fluides,
  nous ne pouvons douter que les particules qui les composent ne soient
  mat\'erielles, et que par cette raison les lois g\'en\'erales de
  l'\'equilibre ne leur conviennent comme aux corps solides}\LAB{c16}, \*

\0his conception of Mechanics as based on a variational principle was
adopted universally by the physicists that developed further the atomistic
theories: in particular Clausius and Boltzmann make constant use of the
action principle using also Lagrange's notations, \cite[p. 25]{Bo866},
\cite{Cl871}, including the commutation rule $\d d=d\d$ (simply reflecting
the equality of the second derivatives with respect to different arguments)
become universal after Lagrange introduced it in the calculus of
variations, \cite[T.I, p.337]{La867}.

In this respect it is worth mentioning that by introducing the variation
$\d$ of the entire extremal curves rather than relying on Euler's local
variations (changing the curves only in infinitesimally close points)
answered in a simple way Euler's question, \cite[T. I, p.336]{La867}, on
why his analysis led to replacing $pdV$ by $-Vdp$ in his calculations:
although simple it was a major change in point of view leading to the
modern formulation of the calculus of variations.

\0(6) Very often the ``Lagrange multipliers'', that he introduced,
\cite[T.I, p.247]{La867}, (extin various cases calculus of variations problems,
are incorporated in the variational equation of Euler-Lagran\-ge, see also
\cite[T.XI, p.340]{La867}.

\0(7) It is important to comment also on the style of all his papers which
continues to inform the {\it M\'ecanique Analytique}: at a time when
quoting other works was not very usual, all papers of Lagrange start with a
dense summary of the previous works on the subject from which he draws the
foundations of his contributions and gives an important help to the
historians of science.

\0(8) The terminology used by Lagrange has become the adopted terminology
in most cases: at times, however, the reader might be confused by the use
of terms which no longer have the same meaning and sorting things out
requires very careful reading: an example is in \cite[T.XI, p.244]{La867}
where the footnote added by J.  Bertrand provides an essential help to the
reader who has not read enough in detail the earlier pages and tomes.

\0(9) In the {\it M\'ecanique Analytique} the theory of the small
oscillations in systems of several degrees of freedom is treated as a
perturbation theory problem and applied to the rigid mody motions and to
celestial mechanics: it was employed extensively in the 1800-ths, starting
with Laplace, until Poincar\'e made clear the need of substantial
improvements on the method by pointing out the existence of motions that
could not be reduced to simple ``quasi periodic'' ones. New ideas came much
later with the work of Siegel, Kolmogorov, Arnold, Moser, Eliasson:
Lagrange's imprint on the problem however remains not only through the
secular perturbations treatment but also through another of his
contributions, namely the theory of the implict functions, Eq.\equ{e3.2},
as outlined in the next section.

\def\SEC{KAM}\section{\SEC}
\label{sec5}\iniz

The problem of stability of quasi periodic motions is closely related to
Lagrange's inversion formula Eq.\equ{e3.2} and to its graphical version
\equ{e3.4}. It can be formulated as an implicit functions problem for a
function $\V h(\Ba)$ defined on the torus $T^\ell$ satisfying, in the
simplest non trivial cases, the {\it Hamilton--Jacobi} equation
\be {\cal K}\V h(\Ba)\defi \V h(\Ba)+(\Bo\cdot\BDpr)^{-2}\Big(\e\BDpr
f(\Ba+\V h(\Ba))\Big)=\V 0\Eq{e5.1}\ee
where $\Bo\in R^\ell$ is a vector with Diophantine property:
$|\Bo\cdot\Bn|\ge C_0^{-1}|\n|^{-\t}$ for some $C_0,\t>0$ and $\BDpr f$ is
the gradient of an analytic even (for simplicity) function on $T^\ell$,
$(\Bo\cdot\BDpr)^{-2}$ is the linear (pseudo)differential operator on the
functions analytic and odd on $T^\ell$ applied to the function of $\Ba$
$\Ba\to \e\BDpr f(\Ba+\V h(\Ba))$ and $\V h(\Ba)$ is to be determined, odd
in $\Ba\in T^\ell$.

This is leads to consider an infinite dimensional version of Lagrange's
inversion: it can be solved in exactly the same way writing $\V A=\V
h+{\cal K}\V h$ and solving for $\V h$ when $\V A=\V 0$, as in
Eq.\equ{e3.2}.  This is more easily done if the Eq.\equ{e5.1} is considered
in Fourier's transfom. Writing $\V h(\Ba)=\sum_{k=1}^\infty \e^k \V
h^{[k]}(\Ba)$ and denoting $\V h_{\Bn,j}^{(k)}$ the Fourier transform of 
$\V h^{(k)}_j,\,\Bn\in Z^\ell$ 
an expression for $j$-th component $h^{[k]}_{\Bn,j}$ is given
via trees with $k$ root-oriented branches as:

\eqfig{300}{140}{
\ins{-5}{71}{$r$}
\ins{-5}{91}{$j$}
\ins{14}{71}{$\Bn\kern-2pt=\kern-2pt\Bn_{\l_{0}}$}
\ins{24}{91}{$\l_{0}$}
\ins{50}{71}{$v_0$}
\ins{46}{91}{$\Bn_{v_0}$}
\ins{127}{100.}{$v_1$}
\ins{121.}{125.}{$\Bn_{v_1}$}
\ins{92.}{41.6}{$v_2$}
\ins{158.}{83.}{$v_3$}
\ins{191.7}{133.3}{$v_5$}
\ins{191.7}{100.}{$v_6$}
\ins{191.7}{71.}{$v_7$}
\ins{210}{5}{$v_{11}$}
\ins{141.7}{20}{$v_{12}$}
\ins{191.7}{18.6}{$v_{10}$}
\ins{166.6}{42.}{$v_4$}
\ins{191.7}{54.2}{$v_8$}
\ins{191.7}{37.5}{$v_9$}
}{Fig2}{Fig.3}
\*
\0Fig.3: {\small A tree $\th$ con
$k_{v_0}=2,k_{v_1}=2,\ldots$ and
$k=13$, with a few decorations\vfil}

In this case on each node there is an extra label $\Bn_v$ (marked in Fig.3
only on $v_0$ and $v_1$) and on each line there is an extra label
$\Bn(\l)=\sum_{w\le v} \Bn_w$; also in this case the labels $j_v,
j_{v_1},\ldots,j_{v_{k_v}}$ associated with the nodes have to be contracted
when appearing twice (\Ie unless $v=r$ as $j_r=j$ appears only once). The
trees will be identified if reducible to each other by pivoting as in the
simple scalar case of Sec.\ref{sec3}. After some algebra it appears that

\be\eqalign{\V h^{[k]}_\Bn=&\sum_\th {\rm Val}(\th),\qquad
{\rm Val}(\th)=\Big(\prod_v \frac{f_{\Bn_v}}{k_v!}\Big)\Big(\prod_\l
\frac{\Bn_v\cdot\Bn_{v'}}{(\Bo\cdot\Bn(\l))^2}\Big)\cr}\Eq{e5.2}
\ee
where the sum is over all trees with $k$ branches.

Eq.\equ{e5.2} was developed in the context of celestial mechanics by
Lindstedt and Newcomb. It turns the proof of the KAM theorem into a simple
algebraic check in which the main difficulty of the {\it small divisors},
which appears because a naive estimate of $\V h^{[k]}_\Bn$ has size
$O(k!^\t)$ (making the formula illusory, because apparently divergent), can
be solved by checking that the values of the trees which are too large have
competing signs and almost cancel between themselves leading to an estimate 
 $|\V h^{[k]}_\Bn|\le c^k e^{-\k |\Bn|}$ for suitable $c,\k>0$.
 
The tree representation is particularly apt to exhibit the cancellations
that occur: the consequent proof of the KAM theorem, \cite{Ga994b}, is not
the classical one and it is often considered too complicated. In this
respect a comment of Lagrange is relevant: \*

\0{\it D'ailleurs mes recherches n'ont rien de commun avec le
  leurs que le probl\`eme qui en fait l'object; et c'est toujours
  contribuer \`a l'avancement des Math\'e\-matiques que de montrer comment
  on peut r\'esoudre les m\^eme questions et parvenir au m\^eme resultats
  par des voies tr\`es-diff\'erentes; les m\'ethodes se pr\`etent par ce
  moyen un jour mutuel et en acqui\`erent souvent un plus grand degr\'e
  d'\'evidence et de g\'en\'eralit\'e.}\LAB{c17} 
\*

\0\cite[T. VI, p.280]{La867}.

Going back to Kepler's problem the work of Carlini and of Levi-Civita
(independent, later) made clear that Lagrange's series, Eq.\equ{e3.6}, can
be resummed into a power series in the parameter $\h=\frac{
  e\,\exp\sqrt{1-e^2}}{1+\sqrt{1-e^2}}$ with radius of convergence $1$,
\cite[Appendice, p.44]{Ca818},\cite{LC904}, thus redetermining the
D'Alembert's radius of convergence $r^*=0.6627434...$ of the power series
in $e$ (called ``Laplace's limit'')\footnote{This is not the only
  resummation of Lagrange's series: the most famous is perhaps Carlini's
  resummation in terms of the Bessel functions $J_n(z)$, namely $\x=\ell
  +\sum_{n=1}^\infty \frac2n J_n(n e) \sin n\ell$, found at the same time
  by Bessel, \cite{Ca817}, \cite{Ca818}, \cite{Be818}, \cite{Ja850},
  \cite{Co992}.} as the closest point to $0$ of the curve $z=e(\h),|\h|=1$,
\Ie $e=i r^*$ or
$r^*\frac{\exp(\sqrt{1+r^{*2}})}{1+\sqrt{1+r^{*2}}}=1$. Furthermore for $e$
real and $e<1$ it is $\h<1$ and an expansion for the eccentric anomaly is
obtained by a power series in $\h(e)$ convergent for all eccentricities
$e<1$.

Recently the same formula has been used to study resonant quasi periodic
motions in integrable systems subject to a perturbing potential $\e V$ with
$\e$ small: the resulting tree expansion allowed the study of the series in
cases in which it is likely to be not convergent: in spite of this it has
been shown that a resummation is possible and leads to a representation of
the invariant torus which is analytic in $\e$ in a region of the form,
\cite{GG005e},

\eqfig{250}{110}{ \ins{75}{95}{$ \rm complex \atop \e\rm-plane$}
}{Fig4}{Fig.4} \*\* 

\0where the contact points of the holomorphy region and
the negative real axis have a Lebesgue density point at $0$; and in other
cases to a fractional power series representation, \cite{GGG006}, or also
to a Borel summability in a region with $\e=0$ on its boundary,
\cite{CGGG006}.

The resummation leading to the above result is a typical resummation that
appears in the eighteenth century analysis and is precisely the one that is
used against the objections of D'Alembert to the vibrating string solution
(\eg $\sum_{k=0}^\infty x^k=(1-x)^{-1}$ for $x>1$): see the quotation
following Eq.\equ{e5.1} above:
%
the modernity
of Lagrange's defying viewpoint will not escape the readers' attention.  \*\*

\def\SEC{Comments}
\section{\SEC}
\label{sec6}\iniz

\0(1) The work of Lagrange has raised a large number of comments and deep
critical analysis, here I mention a few: \cite{Fr983}, \cite{Fr985},
\cite{Pa003}, \cite{AFG004}.
\\
(2) Although born and raised in Torino there are very few notes written in
Italian: all of them seem to be in mail exchanges; a remarkable one has
been inserted in the Tome VII of the collected works, \cite[T.VII,
  p. 583]{La867}. It is a very deferent letter in which he explains a
somewhat unusual algorithm to evaluate derivatives and integrals. Consider
the series

\be [xy]^m\defi \sum_{k=0}^\infty{m\choose k} [x]^{m-k}[y]^k\Eq{e6.1}\ee
(which the 
Lagrange writes without the brackets, added here for clarity). The
positive powers are interpreted as derivatives (or more properly as
infinitesimal increments with respect to the variation of an {\it unspecified}
variable) and the negative as integrals provided at least $x$ or $y$ is an
infinitesimal increment. Thus for $m>0$ and integer the expression is a
finite sum and yields the Leibnitz differentiation rule
$[xy]^m=\sum_{k=0}^\infty{m\choose k} (dx)^{m-k}(dy)^k$. If $m=-1-k<0$
then $[dx]^{-1-k}$ is interpreted as $k+1$ iteration of indefinite
integration over an infinitesimal interval $dx$; this means $dx^{-1}=x,
dx^{-2}=\frac{x^2}{2dx},\ldots, [dx]^{-1-k} =\frac{x^{k+1}}{(k+1)!\,
  dx^k}$.
 
For $m=-1$ Lagrange gives the example $\int y dx$: from Eq. \equ{e6.1}

\be \eqalign{ &[dx\, y]^{-1}=\sum_{k=0}^\infty (-1)^k [dx]^{-1-k} d^ky\cr
  &\int y dx= \sum_{k=0}^\infty \frac{(-1)^k x^{k+1}}{(k+1)! dx^k} d^ky=
\sum_{k=0}^\infty \frac{(-1)^k x^{k+1}}{(k+1)!}\frac{d^ky}{dx^k}
\Eq{e6.2}\cr} \ee
(a relation that the 18 years old Lagrange attributes to ``Giovanni
Bernoullio'', 1694). A further example is worked out:

\be [dx dy]^{-2}=\sum_{k=0}^\infty {-2\choose k} [dx]^{-2-k} d^{k+1}y
=\sum_{k=0}^\infty(-1)^k\frac{k+1}{(k+2)!} x^{k+2}\frac{d^{k+1}y}{dx^{k+1}}
\Eq{e6.3}\ee
which therefore yields the indefinite integral $\int\int dy\,dx$ which is
shown to be identical to $\int y dx$ simply by differentiating the {\it
  r.h.s.} of \equ{e6.3} and checking its identity with the {\it r.h.s.} of
\equ{e6.2} (Lagrange suggests, equivalently, to differentiate both
equations twice).

The letter is signed {\it Luigi De La Grange}, and
addressed to ``Illustrissimo Signor Da Fagnano'', well known mathematician,
who would soon help Lagrange to get his first paper published.
\\ 
(3) Consideration of friction is not frequent in the works of Lagrange, it
is mentioned in a remark on the vibrating strings theory, \cite[T.I, p.109,
  241]{La867}, and the analysis on the tautochrone curves in T.II,III. 

It is also considered in the astronomical problems to examine the
consequences, on the variations of the planetary elements, of a small
rarefied medium filling the solar system (if any) in the Tomes VI and VII.
Or in the influence of friction on the oscillations of a pendulum in the
M\'ecanique Analytique (Tome XII) and in fluid mechanics problems.

Although his attention to applications has been constant (for instance
studying the best shape to give to a column to strengthen it) friction
enters only very marginally in the remarkable theory of the anchor
escapement, \cite[T.IV, p.341]{La867}: this is surprising because it is an
essential feature controlling the precision of the clocks and the very
possibility of building them, \cite[Ch.1,Sec.2.17]{Ga985b}.
\\
(4) Among other applications discussed by Lagrange are problems in Optics,
again referring to the variational properties of light paths, and in
Probability theory.  
\\
(5) Infinitesimals, in the sense of Leibnitz, are pervasive in his work (a
nice example is the 1754 letter to G.C. Fagnano quoted above which shows
that Lagrange learnt very early their use and their formidable power of
easing the task of long algebraic steps (at the time there were already
some objections to their use): this makes reading his papers easy and
pleasant; at the time there were several objections to their use which
eventually lead to the rigorous refoundation of analysis.

However Lagrange himself seems to have realized that something ought to be
done in systematizing the foundations of analysis: the tellingly long title
of his lecture notes, \cite[T. IX]{La867}, {\it Th\'eorie des Fonctions
  analytiques, contenants les principes du calcul differentiel d\'egag\'e
  de toutes consid\'erations d'infiniment petits ou d'evanouissants, de
  limites ou de fluxions et r\'eduits \`a l'analyse alg\'ebrique de
  quantit\'es finies}, and the very first page of the subsequent {\it
  Le\c{c}ons sur le calcul des fonctions}\LAB{c18}, \cite[T. X]{La867},
\*
\0{\it On conna\^\i{}t les difficult\'es qu'offre la supposition
des infiniment petits, sur laquelle Leibnitz a fond\'e le Calcul
différentiel. Pour les \'eviter, Euler regarde les diff\'erentielles comme
nulles, ce qui r\'eduit leur rap- rapport \`a l'expression z\'ero divis\'e
par z\'ero, laquelle ne pr\'esente aucune id\'ee.}\LAB{c19}
\*
are a clear sign of his hidden qualms on the matter.

The fact that they developed at a late stage, while he
wholeheartedly adopted the Leibnitz methods in his youth, shows that the
problem of mathematical rigor had grown, even for the great scientists, to
a point that it was necessary to work more on it. Nevertheless his last
words on the subject are probably to be found in the preface to the second
edition of the {\it M\'ecanique Analytique} where he (reassuringly) admits
\*
\0{\it On a conserv\'e la notation ordinaire du Calcul diff\'erentiel,
  parce qu'elle r\'epond au syst\`eme des infiniment petits, adopt\'e dans
  ce Trait\'e.  Lorsqu'on a bien con\c{c}u l'esprit de ce syst\`eme, et
  qu'on s'est convaincu de l'exactitude de ses r\'esultats par la m\'ethode
  g\'eom\'etrique des premi\`eres et derni\`eres raisons, ou par la
  m\'ethode analytique des fonctions d\'eriv\'ees, on peut employer les
  infiniment petits comme un instrument s\^ur et commode pour abr\'eger et
  simplifier les d\'emonstrations. C'est ainsi qu'on abr\`ege les
  d\'emon\-strations des Anciens par la m\'ethode des indivisibles.
}\LAB{c20}  \\
(6) An informative history of his life can be found in the ``Notices sur la
vie et les ouvrages'' in the preface by M. Delambre of the Tome I of the
collected works, \cite{La867} and in the commemoration by P. Cossali at the
University of Padova, \cite{Co813}. If abstraction is made of the bombastic
rethoric celebrating, in the first (quite a) few pages, Eugene Napoleon
and, in the last (quite a) few pages, Napoleon I himself the remaining
about 130 pages of Cossali's {\it Elogio} contain a useful and detailed
summary and evaluation of all the works of Lagrange (exposed, as well, in a
circumvoluted rethorical style).
(7) A recent analysis of Lagrange's contributions to Celestial Mechanics in
perspective with the historical development up to contemporary works can be
found in \cite{Fe013}.

\def\SEC{References}

\small
\bibliography{0Bib}


\centerline{\bf End Notes: translation of the quotations}
\*

\parindent=0mm
\*\raf{c1}
%
From the above considerations follows that the analysis that we have
proposed in the previous chapter is perhaps the only one which could thorow
on such obscure subjects a light sufficient to clarify the doubts arising
from various sources

\*\raf{c2}
%
I do not believe that it for this never has been given a genaral formula,
like the one that we just gave

\*\raf{c3}
%
But I ask him to please pay attention that, in my solution, the
determination of the string shape at each instant depends uniquely on the
quantities $Z$ and $U$, which do not enter, by any means, in the considered
operation. I agree that the formula to which I apply Mr.  Bernoulli's
method is subject to the continuity law; however it does not seem to me
that the quantities $Z$ and $U$, constituting the coefficients of this
formula, are also subject to it, as D'Alembert pretends.

\*\raf{c4}
%
Now I ask whether every time that in an algebraic formula it eill for
instance occur an infinite geometric series, such as $1 +x +
x^2+x^3+x^4+\ldots$, one had not the right to replace it by $\frac1{1-x}$
although this quantity is not really equal to the sum of the proposed
series unless one supposed that the last term $x^\infty$ vanishes.  It
seems to me that we could not context the correctness of such a
substitution without overthrowing the most common principles of analysis.

\*\raf{c5}
%
But, one would say, how can it be that the sum of the infinite sequence
$\cos x+ \cos2x+ \cos3x+...$  is always equal to $-\frac12$ since, in the
case $x=0$, it becomes necessarily equal to a sequence of as many unities?
I answer that ...

\*\raf{c6}
%
I answer that by a similar argument one would also maintain that
$\frac1{1+x}$ is not the general expression of the sum of the infinite
sequence $1-x+x^2-x^3+\ldots$ because, setting $x=1$, one gets
$1-1+1-1+\ldots$ which is eithe $0$ or $1$ depending on whether the number
of terms considered is even or odd, while the value of $\frac1{1+x}$ is
$\frac12$. Now I do not believe that any Geometer would be willing to admit
such conclusion

\*\raf{c7}
%
I admit that I did not express myself sufficiently exactly

\*\raf{c8}
%
I confess the that they do not look to me strong enough to infirm your
solution. The great genius seems to me too incline to destroy what he
himself does not conctruct.\\ ...\\ After this remark, I easily concede,
Sir, that in order that the string motion be conform to the continuity law,
it is necessary that the initial shape the derivatives $\frac{d^2
y}{dx^2},\frac{d^4 y}{dx^4}, \frac{d^6 y}{dx^6}$ be equal to $0$ at the
extremes but, whether or not such conditions take place, I believe that I
could maintain that our solution will neverthesless give the true motion of
the string because ...

\*\raf{c9}
%
The transformations that I use here are those called integration by parts,
and which are normally proved by the principles of differential calculus;
but it is not difficult to see that their foundations lie in the general
calculus of summs and differences; hence it follows that one does not at
all have to fear that ffor this any continuity law between the different
$z$-values is introduced in our calculation

\*\raf{c10}
Such equations, as is easy to see, equal un number that of the particues of
which we study the motions; therefore the probem being already fully
determined via them, one is obliged to follow course, so that any foreing
condition cannot fail to make the solution not sufficient and possibly even
wrong.

\*\raf{c11}
%
Decompose each of the motions $a,b,c\&tc$ impressed to each body in two
others $\a,\a';\b,\b';\g,\g'\&tc$ such that if we impressed on the bodies
only the motions $\a,\b,\g,\&tc$ that they could have kept such motions
without influencing each other; and if we had not impressed any motion
other than the motions $\a',\b',\g',\&tc$ the system would have remained
motionless; it is clear that $\a,\b,\g,\&tc$ will be the motions that the
bodies will undergo becuse of their actions. WHat was to be found.

\*\raf{c12}
%
.. That is just the dynamical principle of M. d'Alembert, transformed into
a formula via the law of virtual velocities. But the combination of such two
principles is a step which had never been done, and thisis erhaps the only
degree of perfection which, after M. d'Alembert's discovery, was still
missing in the Theory of Dynamics

\*\raf{c13}
%
%
For motions of an arbitrary system driven by forces mutually attractive, or
directed towards fixed centers and proportional to arbitrary functions of
the distances, the curves described by the various bodys, as well as their
velocities, are necessarily such that the velocity times the line element
of the curve is maximum or a minimum, provided the first and last point of
each curve are regarded as given, so that the variations of the
corresponding coordinates vanish.
\\
This is the theorem about which we spoke at the end of the first section,
under the name of Principle of least action,

\*\raf{c14}
%
... this demands the resolution of a cubic equation. However, thinking
about the essence of the problem, it seems that it should be possible to
solve it directly and independentlyof the rotation axes properties, whose
demonstration is rather difficult and which, on the other hand, should
rather be consequences of the solution itself than of its the foundation.

\*\raf{c15}
%
This is the case in which the coordinates axis $c$, \Ie the line passing
through the suspension point and the baricenter is a natural rotation axis,
and in which the inertia moments around the other two axes are equal
(art. 32), this happens in general for ll solids of rotation when the fixed
point is on the revolution axis. The solution of this case ie easy because
of the three integrals just found.

\*\raf{c16}
%
Although we ignore the internal constitution of fluids, we cannot doubt that
the particles that compose them are material, and for this reason the the 
general equilibrium laws pertain to them as they do to solid bodies

\*\raf{c17}
%
However my researches have nothing else in common with theirs besides the
problem of which they are the object: and it is always a contribution to
the progress of Mathematiques to show how the same questions can be solved
and the same results obtained through very different ways; the methods
provide in this was mutual support and often a better degree of evidence
and generality.

\*\raf{c18}
%
Analytic functions theory, containing the principles of differential
calculus freed of all considerations of infinitesimal or evanescent
quantities, of limits or fluxions and reduced the the algebraic analysis of
finite quantities

\*\raf{c19}
%
We know the difficulties arising when assuming the infinitesimals, on
which Leibnitz built differential calculus. To avoid them Euler condiders
the differentials as vanishing, which reduces their ratios to the
expression  zero divided by zero, which does not suggest any idea ..

\*\raf{c20}
%
We kept the ordinary differential Calculus notation, because it agrees with
the syste of infinitesimals, adopted n this Trait\'e. When the spirit of
this systme has been well understood, and one is convinced of the exactness
of the results through the geometrical method of the first and last ratios,
or through the analytical method of the derivative functions, one can use
the infinitesimals as a sure and convenient instrument to abridge and
simplify the proofs. It is in this way that the classical proofs are
abridged via the method of the indivisibles.

%
%
\end{document}